\documentclass[aps,prb,reprint,superscriptaddress]{revtex4-2}

\usepackage{color}
\usepackage{graphicx}
\usepackage{dcolumn}
\usepackage{bm}
\usepackage{hyperref}
\hypersetup{colorlinks=true,linkcolor=blue,citecolor=blue,urlcolor=blue}

\usepackage[T1]{fontenc} 
\usepackage{newtxtext}   
\usepackage{newtxmath}   
\usepackage{upgreek}
\usepackage{mhchem}
\usepackage{tabularx}
\usepackage{textcomp}

\begin{document}

\preprint{APS/123-QED}

\title{Phase calibration of quantum oscillations in the magnetostrictive coefficient using the topological antiferromagnet \ce{YbMnBi2}}  

\author{Qin Deng}
\thanks{These authors contributed equally to this work. }
\affiliation{Low Temperature Physics Laboratory, School of Physics \& Center of Quantum Materials and Devices, Chongqing University, Chongqing 401331, China}

\author{Long Zhang}
\thanks{These authors contributed equally to this work. }
\affiliation{Low Temperature Physics Laboratory, School of Physics \& Center of Quantum Materials and Devices, Chongqing University, Chongqing 401331, China}
\affiliation{College of Materials Science and Engineering \& Center of Quantum Materials and Devices,  Chongqing University, Chongqing 401331, China}

\author{Zeyu Li}
\thanks{These authors contributed equally to this work. }
\affiliation{Low Temperature Physics Laboratory, School of Physics \& Center of Quantum Materials and Devices, Chongqing University, Chongqing 401331, China}

\author{Ying Zhu}
\thanks{These authors contributed equally to this work. }
\affiliation{Low Temperature Physics Laboratory, School of Physics \& Center of Quantum Materials and Devices, Chongqing University, Chongqing 401331, China}

\author{Shuai Wu}
\affiliation{Low Temperature Physics Laboratory, School of Physics \& Center of Quantum Materials and Devices, Chongqing University, Chongqing 401331, China}

\author{Yan Liu}
\affiliation{Analytical and Testing Center of Chongqing University, Chongqing University, Chongqing 401331, China}

\author{Aifeng Wang}
\affiliation{Low Temperature Physics Laboratory, School of Physics \& Center of Quantum Materials and Devices, Chongqing University, Chongqing 401331, China}

\author{Yu Pan}
\email{yupan2024@cqu.edu.cn}
\affiliation{College of Materials Science and Engineering \& Center of Quantum Materials and Devices,  Chongqing University, Chongqing 401331, China}
\affiliation{Mingyue Lake Laboratory, Chongqing 401135, China}

\author{Yisheng Chai}
\email{yschai@cqu.edu.cn}
\affiliation{Low Temperature Physics Laboratory, School of Physics \& Center of Quantum Materials and Devices, Chongqing University, Chongqing 401331, China}

\date{\today}

\begin{abstract}
The Berry phase accumulated along a cyclotron orbit encodes important information about electronic band topology and is commonly inferred from the phase of quantum oscillations. Measurements of the ac magnetostrictive coefficient have recently emerged as a sensitive thermodynamic probe of quantum oscillations, but the phase offset has not been experimentally calibrated. Here, using the topological antiferromagnet \ce{YbMnBi2}, we calibrate this offset by directly comparing quantum oscillations in magnetization with those in the ac magnetostrictive coefficient. Measurements of both responses on the same single crystal reveal a single fundamental frequency of approximately 160 T in fields up to 14 T, enabling a direct phase comparison free from ambiguities associated with multiple frequencies. We observe an approximately $\pi/2$ relative phase shift between the two oscillatory responses, consistent with the Maxwell relation linking the magnetostrictive coefficient to the stress derivative of magnetization. Our results establish the appropriate phase needed to extract cyclotron-orbit phase information from quantum oscillations in the ac magnetostrictive coefficient.
\end{abstract}

\maketitle

\section{Introduction}

The emergence of a nontrivial Berry phase in the electronic structure near the Fermi energy can give rise to intriguing properties in topological materials, including the anomalous Hall effect and the chiral anomaly \cite{Hasan2010,xiaoliang2011,Ren_2016}. Determining the Berry phase has therefore become an important topic in modern condensed matter physics. Experimentally, cyclotron-orbit phase information is commonly inferred from quantum oscillations in electrical transport (the Shubnikov--de Haas effect), thermoelectric transport, magnetization (the de Haas--van Alphen effect), and magnetic torque \cite{Wright_QO,Alexandradinata_QO,Zhao_QO,FOMINYKH_QO,Hu_QO}. Quantum oscillations also provide information about Fermi-surface geometry and quasiparticle effective masses. They thus offer a powerful route to the electronic structure governing the physical properties of metals.

In addition to these commonly used probes, magnetostriction provides another sensitive channel for detecting quantum oscillations \cite{Chandrasekhar_MS,Eremenko_MS,Budko_MS,FINKELSTEIN_MS,Chandrasekhar_MS2}. The magnetostriction or magnetostrictive strain $\epsilon_i$, where $i=1,\ldots,6$ in contracted notation, quantifies the strain induced by an external magnetic field $B$. Like magnetization $M$, magnetostriction is a thermodynamic quantity and therefore probes the bulk electronic structure. Within Lifshitz--Kosevich (LK) theory, the leading rapidly oscillating contributions to magnetization and magnetostriction are (see the Appendix for details): \cite{LK1956,Chandrasekhar_MS,Hu_QO}
\begin{equation}
    M^\mathrm{osc}=-\sum_{\substack{r=1 }}^\infty A_r\frac{2\pi rF}{B^2}\sin\left(\frac{2\pi rF}{B}-\Phi_r\right),
    \label{eq:M-leading}
\end{equation}
\begin{equation}
    \epsilon^\mathrm{osc}_i=\sum_{\substack{r=1 }}^\infty A_r\frac{2\pi rF}{B}\frac{\partial \ln F}{\partial \sigma_i}\sin\left(\frac{2\pi rF}{B}-\Phi_r\right),
    \label{eq:epsilon-leading}
\end{equation}
where $A_r$ is the slowly varying amplitude of the oscillatory thermodynamic potential, $r$ is the harmonic index,  and $\sigma_i$ is the stress. The fundamental frequency $F$ is related to the extremal Fermi-surface cross-sectional area $S_m$ perpendicular to the magnetic field through the Onsager relation $F=\hbar S_m/(2\pi e)$. For a three-dimensional (3D) extremal orbit, the phase may be written as $\Phi_r=2\pi r\gamma-2\pi\delta$, where $\delta=\pm1/8$ is the curvature correction. In the conventional spin-degenerate LK description, $\gamma=1/2-(\phi_B+\phi_R)/(2\pi)$ contains the Berry phase $\phi_B$ and the orbital-moment (Roth) phase $\phi_R$, whereas the Zeeman contribution is represented separately by the spin factor $R_\mathrm{S}$ [see Eq. \ref{eq:RS}] \cite{Alexandradinata_QO,FOMINYKH_QO}. Equations~(\ref{eq:M-leading}) and (\ref{eq:epsilon-leading}) show that oscillating magnetization and magnetostriction are either in phase or antiphase, depending on the sign of $\partial\ln F/\partial\sigma_i\propto\partial\ln S_m/\partial\sigma_i$. Comparing the oscillating amplitudes of magnetization and magnetostriction can therefore reveal the stress dependence of $S_m$, as demonstrated in several metals \cite{Chandrasekhar_MS,Eremenko_MS,Budko_MS,FINKELSTEIN_MS,Chandrasekhar_MS2}. Conversely, oscillations in magnetostriction vanish to leading order when the extremal orbit is insensitive to stress.

When the Fermi surface is stress sensitive, the magnetostrictive coefficient $\partial\epsilon_i/\partial B$ can be even more sensitive to quantum oscillations \cite{chai2024_QO,chai2026_QO}. Differentiating Eq.~(\ref{eq:epsilon-leading}) and retaining the dominant derivative of the phase gives:
\begin{equation}
    \frac{\partial\epsilon^\mathrm{osc}_i}{\partial B}=-\sum_{\substack{r=1 }}^\infty A_r\frac{(2\pi rF)^2}{B^3}\frac{\partial \ln F}{\partial \sigma_i}\sin\left(\frac{2\pi rF}{B}-\Phi_r+\frac{\pi}{2}\right),
    \label{eq:lambda-leading}
\end{equation}
The same results follow from the thermodynamic Maxwell relation \cite{Chandrasekhar_MS}:
\begin{equation}
   \left(\frac{\partial\epsilon^\mathrm{osc}_i}{\partial B}\right)_{\sigma_i}=\left(\frac{\partial M^\mathrm{osc}}{\partial \sigma_i}\right)_{B}.
     \label{eq:Maxwell}
\end{equation}
Relative to magnetization and magnetostriction, respectively, the oscillation amplitude of the magnetostrictive coefficient gains additional factors of $2\pi rF/B$ and $2\pi rF/B^2$. This factor favors the detection of high-frequency oscillations at relatively low fields. Indeed, in our previous study of the topological nodal-line semimetal ZrSiS, oscillations in the magnetostrictive coefficient appeared at a lower onset field than those in magnetization and magnetostriction, and additional high-frequency branches were resolved \cite{chai2024_QO}. This sensitivity cannot be reproduced simply by numerically differentiating a less-sensitive magnetostriction trace. Instead, we use a composite magnetoelectric (ME) technique to directly detect the ac magnetostrictive coefficient (see Methods) \cite{chai2024_QO,chai2026_QO,Chai2021_ME,zhang2023_ME,Mi2025_ME}.

Despite this high sensitivity, the response-specific phase of the ac magnetostrictive coefficient has not yet been experimentally calibrated. Equations~(\ref{eq:M-leading}) and (\ref{eq:lambda-leading}) predict a relative phase shift of $\pm\pi/2$ between the magnetostrictive coefficient and magnetization. The overall sign depends on the stress derivative of the extremal orbit. Here, we test this theoretical expectation by measuring both quantities on the same single crystal of the topological antiferromagnet \ce{YbMnBi2}, which stands out by displaying only one frequency up to 14 T. Direct LK fits show that their oscillatory components differ in phase by approximately $\pi/2$, which is consistent with the theoretical prediction. This establishes the appropriate phase needed to extract the Berry phase of quantum oscillations measured through the ac magnetostrictive coefficient.

\begin{figure*}
\includegraphics[scale=0.6]{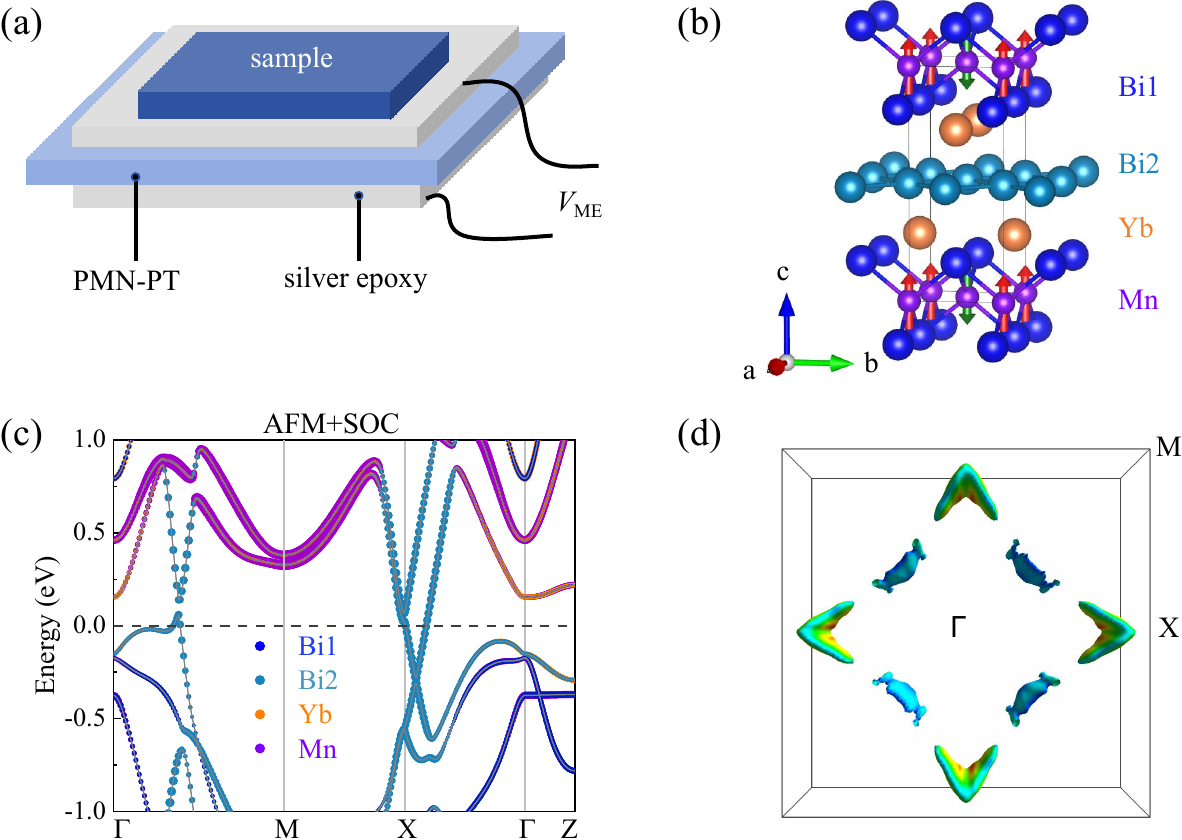}
\caption{\label{fig:1}(a) Schematic setup of the composite ME method used to measure the ac magnetostrictive coefficient. A sample is mechanically bonded to a PMN-PT single crystal with Ag epoxy. In a magnetic field, the magnetostrictive strain $\epsilon_i$ of the sample generates a voltage $V_\mathrm{ME}$ across the PMN-PT through the piezoelectric effect. (b) Side view of the crystal structure of \ce{YbMnBi2}. Arrows on the Mn atoms indicate the magnetic moments in the C-type antiferromagnetic (AFM) structure. (c) and (d) Electronic band structure and Fermi surface of \ce{YbMnBi2} calculated with spin--orbit coupling (SOC) included in the absence of spin canting. Dirac-like bands appear near the Fermi level along the $\Gamma$-- M and $\Gamma$-- X directions. An electronlike pocket around X and a holelike pocket along $\Gamma$--M are visible. }
\end{figure*}

\section{Methods}

The ac magnetostrictive coefficient was measured using the composite magnetoelectric method shown in Fig.~\ref{fig:1}(a) \cite{chai2024_QO,chai2026_QO,Chai2021_ME,zhang2023_ME,Mi2025_ME}. The sample was mechanically coupled with silver epoxy (H20E, EPO-TEK) to a piezoelectric single crystal of 0.7Pb(\ce{Mg1/3Nb2/3})\ce{O3}--0.3\ce{PbTiO3} (PMN-PT). The PMN-PT crystals were prepared as 0.2-mm-thick [001]-cut plates and electrically poled at room temperature under an electric field of 550 kV/m using a Keithley 6517B electrometer. In this configuration, the field-induced in-plane strain of the sample is transferred to the PMN-PT and converted into an electrical voltage $V_\mathrm{ME}$. In this sense, PMN-PT acts as a strain sensor, so that
\begin{equation}
    V_\mathrm{ME} = k\frac{\partial E}{\partial\epsilon_i}\frac{\partial\epsilon_i}{\partial B},
\label{eq:VME}
\end{equation}
where $0<k<1$ represents the strain-transfer efficiency and $\partial E/\partial\epsilon_i$ is the piezoelectric coefficient of PMN-PT. To enhance the sensitivity, an ac magnetic field of 1 Oe generated by a home-made Helmholtz coil was superimposed on the dc field, and the resulting ac voltage $V_\mathrm{ME}$ was detected with a lock-in amplifier. In our setup, the sign of $V_\mathrm{ME}$ was calibrated beforehand, with a positive voltage corresponding to tensile magnetostrictive strain. Because the coupling factor $k$ is not known precisely, the absolute magnetostrictive coefficient cannot be determined from $V_\mathrm{ME}$. Nevertheless, $V_\mathrm{ME}$ closely tracks $(\partial\epsilon_i/\partial B)_\mathrm{ac}$ and therefore preserves its frequency and phase in quantum oscillations.

Single crystals of \ce{YbMnBi2} were grown by a Bi self-flux method, as described in our earlier study \cite{pan2022_YbMnBi}. The crystal used here had dimensions of $2.8\times1.5\times0.2$ mm$^3$. The same crystal was used for the magnetization, magnetostrictive-coefficient and thermopower measurements to minimize sample-dependent uncertainty. Magnetization was measured in a Physical Property Measurement System (PPMS, Quantum Design DynaCool) equipped with a vibrating-sample magnetometer (VSM). The ac magnetostrictive coefficient was measured in a 14-T cryostat (Oxford Instruments). Thermopower was measured in the same cryostat using a one-heater and two-thermometer configuration on a homemade rotator probe.

First-principles calculations were performed within density-functional theory using the projector-augmented-wave method implemented in VASP \cite{Kresse_PBE,Bloch_PBE,Kress_PBE}. Exchange and correlation were treated within the PBE generalized-gradient approximation \cite{Perdew_PBE}, with a plane-wave cutoff energy of 500 eV. The on-site Coulomb interaction of the Mn $3d$ electrons was included using the Dudarev DFT+$U$ method with $U_\mathrm{eff}=3$ eV \cite{Dudarev_U}. The crystal structure was first optimized in a collinear antiferromagnetic configuration without spin--orbit coupling (SOC). Calculations including SOC were then performed for the optimized structure using a $\Gamma$-centered $13\times13\times7$ $k$-point mesh. Band dispersions were evaluated along conventional high-symmetry directions in the Brillouin zone. The Fermi surface was constructed with VASPKIT from eigenvalues calculated on a uniform $28\times28\times12$ $k$-point mesh and visualized using FermiSurfer \cite{WANG2021,KAWAMURA2019}.

\section{Results and Discussion}
A system with a single frequency and no resolved higher harmonics is ideal for phase calibration because it avoids ambiguities associated with overlapping oscillatory components. Our previous electrical-transport study showed that the topological antiferromagnet \ce{YbMnBi2} exhibits one fundamental frequency up to 9 T \cite{pan2022_YbMnBi}. We therefore chose \ce{YbMnBi2} as a model system for calibrating the phase of quantum oscillations in the ac magnetostrictive coefficient measured by the composite ME method.

Figure~\ref{fig:1}(b) shows the crystal structure of \ce{YbMnBi2}, which crystallizes in the tetragonal space group $P4/nmm$ \cite{Borisenko_YbMnBi,Wang_YbMnBi,Liu2017_YbMnBi,pan2022_YbMnBi,Le2021_YbMnBi,Ni2022_YbMnBi}. The Mn--Bi1 layers adopt an anti-PbO-type structure, whereas the Bi2 atoms form square-net layers that dominate the topological electronic structure \cite{Borisenko_YbMnBi,pan2022_YbMnBi,Le2021_YbMnBi,Ni2022_YbMnBi}. This layered arrangement produces a quasi-two-dimensional (2D) electronic structure \cite{Borisenko_YbMnBi,Liu2017_YbMnBi}. Below $T_\mathrm{N}\approx290$ K, the Mn moments align predominantly along the $c$ axis and form a collinear C-type antiferromagnetic (AFM) state, with antiferromagnetic coupling in the $ab$ plane and ferromagnetic coupling along $c$  \cite{Wang_YbMnBi,Borisenko_YbMnBi,Soh_YbMnBi,xie2026_YbMnBi}. No magnetic order of the Yb moments has been reported down to 2 K. The topology of the electronic structure is highly sensitive to the magnetic structure on the Mn sublattice. Theoretically, canting the magnetic moments away from the $c$ axis, thereby generating a finite net ferromagnetic moment, may drive the system from a Dirac semimetal in the collinear state to a type-II Weyl semimetal \cite{Borisenko_YbMnBi,Ni2022_YbMnBi,Le2021_YbMnBi}. The magnetic ground state nevertheless remains under debate. Angle-resolved photoemission spectroscopy, interlayer transport, and optical measurements have been interpreted in terms of a canting angle of approximately $10^\circ$, which induces a time-reversal-symmetry-breaking type-II Weyl state \cite{Borisenko_YbMnBi,Liu2017_YbMnBi,Chinotti_YbMnBi}. By contrast, neutron-scattering experiments have resolved no clear signature of spin canting, suggesting that canting is negligible in bulk \ce{YbMnBi2} \cite{Wang_YbMnBi,Soh_YbMnBi,xie2026_YbMnBi}.

Because magnetization and the magnetostrictive coefficient are bulk probes, we calculated the electronic structure of \ce{YbMnBi2} in the collinear C-type antiferromagnetic configuration without canting. Figures~\ref{fig:1}(c) and \ref{fig:1}(d) show the electronic band structure and  Fermi surface with SOC included. The states near the Fermi level are dominated by Bi $p$ orbitals. Dirac-like bands with nearly linear dispersion occur along the $\Gamma$-- M and $\Gamma$-- X directions. In the absence of spin canting, four Dirac nodes dominated by the Bi2 $p$ orbitals occur within the first Brillouin zone \cite{Borisenko_YbMnBi,Ni2022_YbMnBi}. Spin--orbit coupling and magnetic exchange interactions open gaps at these nodes, producing massive Dirac states. The corresponding bands form an electronlike pocket  near X and a holelike pocket along $\Gamma$-- M. The pockets are well separated, and all bands remain doubly degenerate in the zero-canting state. A canting-induced ferromagnetic component can lift this degeneracy, causing the electronlike and holelike pockets to touch and form type-II Weyl points \cite{Borisenko_YbMnBi,Liu2017_YbMnBi,Ni2022_YbMnBi,Le2021_YbMnBi}.

\begin{figure*}
\includegraphics[scale=0.3]{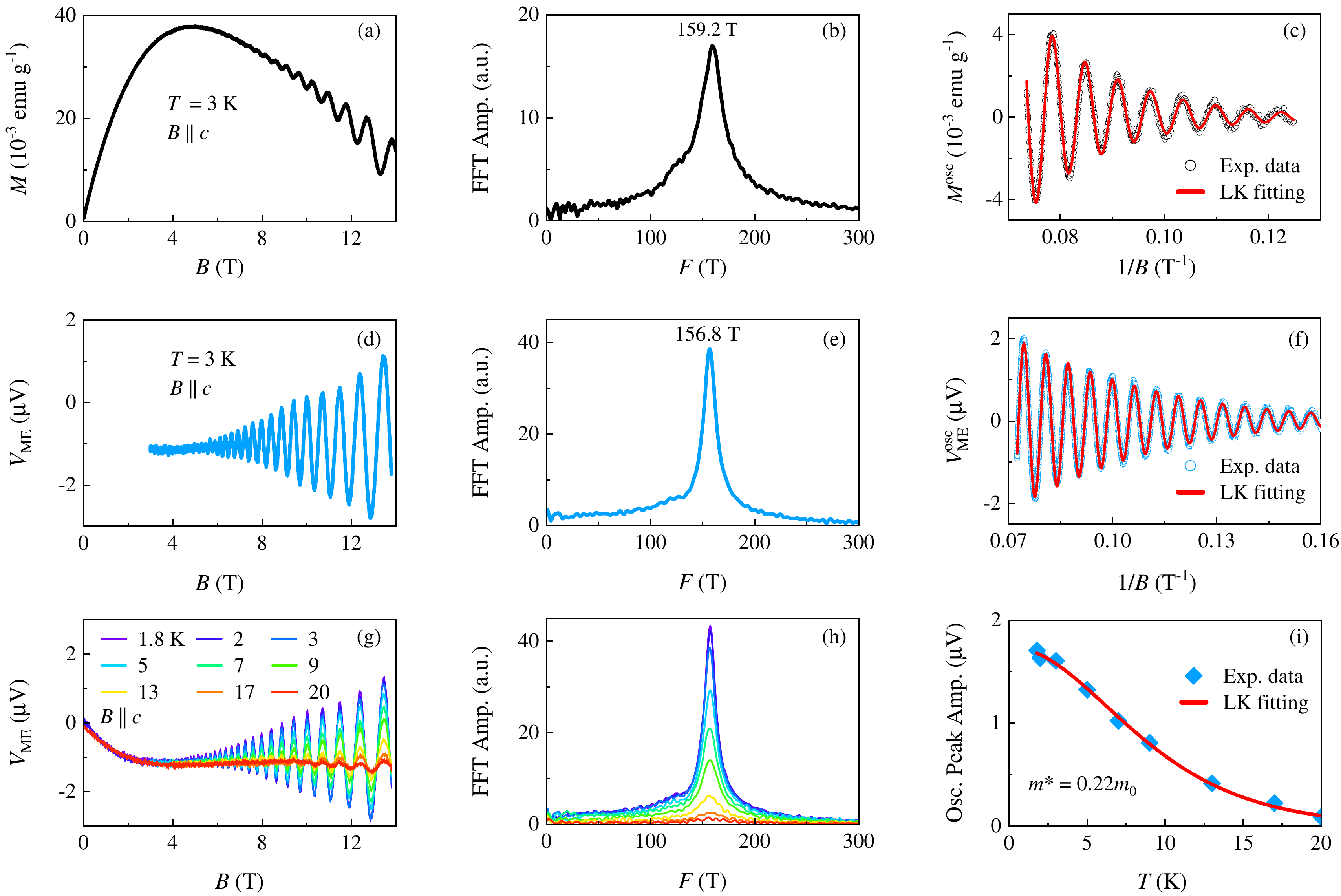}
\caption{\label{fig:2} Quantum oscillations in magnetization and the ac magnetostrictive coefficient for $B\parallel c$. (a,d) Representative raw magnetization and ME-voltage data, with $V_\mathrm{ME}\propto\partial\epsilon_i/\partial B$. (b,e) FFT spectra of the corresponding oscillatory components. (c,f) LK fits (red lines) to the oscillatory parts plotted versus inverse field $1/B$. (g,h) Temperature-dependent oscillations in the ME signal and the corresponding FFT spectra. (i) Fit of the oscillation  peak amplitude at a fixed magnetic field of 12.4 T using the LK thermal damping factor (red line), yielding $m^{*}=0.22m_0$.}
\end{figure*}

Figure~\ref{fig:2} compares quantum oscillations in magnetization $M$ and ME voltage $V_\mathrm{ME}\propto\partial\epsilon_i/\partial B$ measured on the same \ce{YbMnBi2} crystal with $B\parallel c$. Clear oscillations occur in both quantities, but their backgrounds and onset fields differ substantially [see Figs.~\ref{fig:2}(a) and \ref{fig:2}(d)]. The oscillations in magnetization are superimposed on a complex field-dependent background, whereas the background in $V_\mathrm{ME}$ varies only weakly with magnetic field. Oscillations become visible above approximately 4 T in $V_\mathrm{ME}$ but require fields above approximately 6 T in magnetization. The lower onset field confirms the enhanced sensitivity of the magnetostrictive coefficient at low magnetic fields, as also observed in our previous work \cite{chai2024_QO}.

Despite these differences, the two probes resolve similar frequencies.  Fast Fourier transform (FFT) analysis after background subtraction gives single fundamental frequencies of $F=159.2$ T for magnetization and $F=156.8$ T for the ME signal [see Figs.~\ref{fig:2}(b) and \ref{fig:2}(e)]. An earlier magnetotransport study reported a broad dominant frequency near 130 T \cite{Wang_YbMnBi}, whereas measurements up to 45 T resolved two branches, $F_\alpha=115$ T and $F_\beta=162$ T \cite{Liu2017_YbMnBi}. The frequency observed by our bulk probes agrees well with the $F_\beta$ branch, assigned to the larger electronlike pocket near X. This assignment is consistent with the dominance of electronlike carriers in low-field electrical and thermoelectric transport \cite{Wang_YbMnBi,pan2022_YbMnBi}. The small frequency difference between the two measurements most likely arises from a slight field misalignment in this quasi-2D Fermi surface. Later we show that such a small misalignment has negligible effect on the phase of quantum oscillations in \ce{YbMnBi2}.

Using the temperature-dependent $V_\mathrm{ME}$  shown in Fig.~\ref{fig:2}(g), we determine the cyclotron mass $m^{*}$ from the oscillation peak amplitude using the LK thermal factor $R_\mathrm{T}$ [Eq.~(\ref{eq:RT}) in the Appendix].  As shown in Fig.~\ref{fig:2}(i), the temperature dependence of the oscillation amplitude taken at a fixed magnetic field $B=12.4$ T, is well described by $R_\mathrm{T}$ and yields $m^{*}=0.22m_0$, agreeing well with the value $m^{*}=0.24m_0$ obtained from magnetotransport \cite{Wang_YbMnBi,Liu2017_YbMnBi,pan2022_YbMnBi}. Here, $m_0$ is the free electron mass. The small effective mass is consistent with the Fermi pocket derived from Dirac bands.

\begin{figure*}
\includegraphics[scale=0.4]{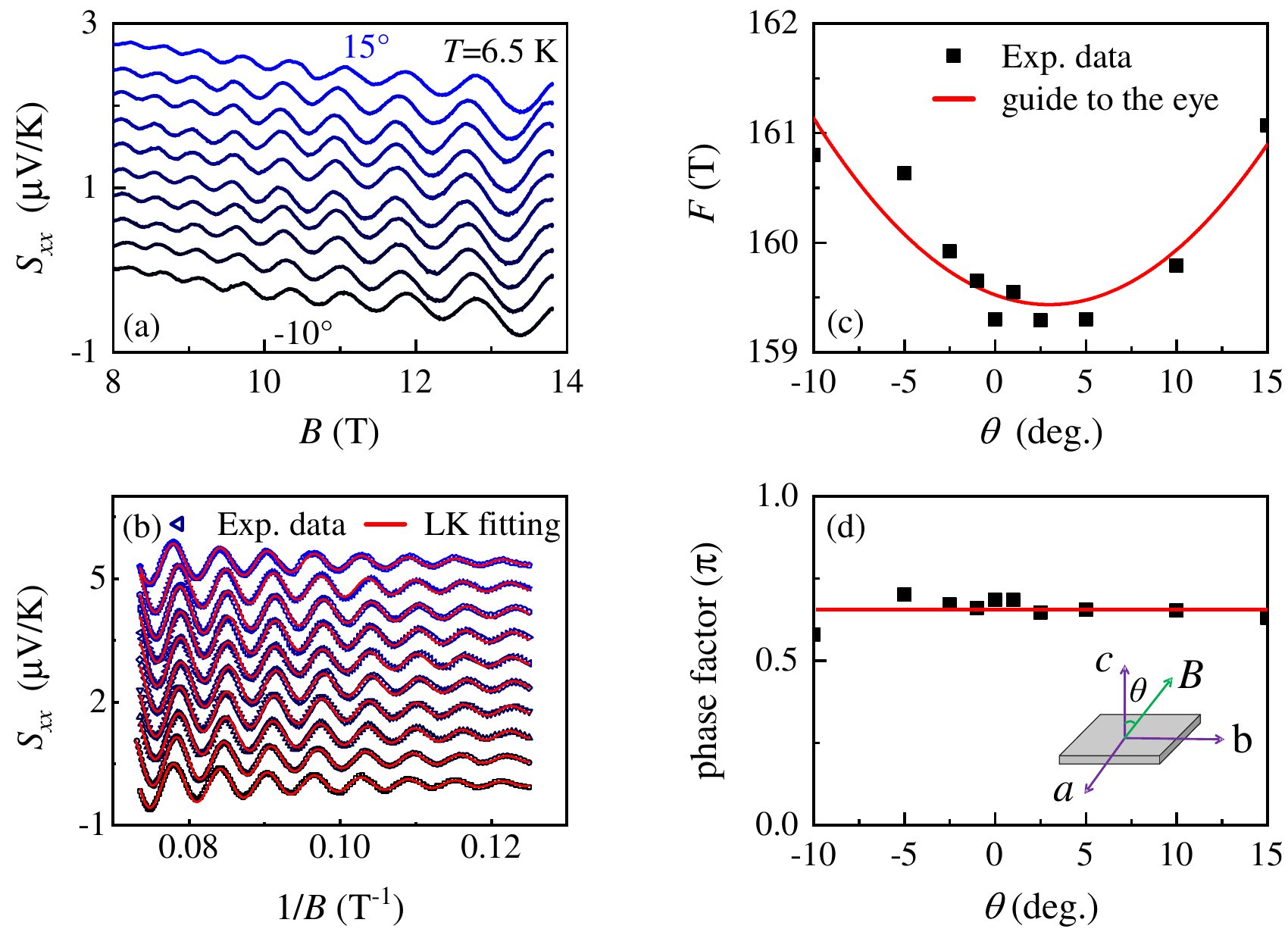}
\caption{\label{fig:3} Quantum oscillations in the thermopower $S_{xx}$ for magnetic fields close to the $c$ axis. (a) $S_{xx}$ measured at selected angles. (b) The oscillator components (open triangles) of $S_{xx}$ fitted with the LK formula (red solid line). All curves in (a,b) except the one at $-10^\circ$ have been vertically offset for clarity.  (c,d) Angular dependence of the fitted oscillation frequency and phase, respectively. Red lines are guides to the eye. The inset in (d) illustrates the measurement geometry and the nominal angle $\theta$ is defined between the magnetic field and the $c$ axis.}
\end{figure*}

\begin{table*}
\centering
\caption{Parameters obtained from LK fits to the oscillatory magnetization and ME voltage. A common positive-amplitude sine convention is used.}
\label{table:1}
\renewcommand*{\arraystretch}{1.5}%
\begin{tabularx}{1\textwidth}{
>{\centering\arraybackslash}X
>{\centering\arraybackslash}X
>{\centering\arraybackslash}X 
>{\centering\arraybackslash}X
>{\centering\arraybackslash}X}
\hline
\hline
   & $C_\mathrm{M}$ or $C_\mathrm{V}$ (a.u.) & $F$ (T) & $\varphi_\mathrm{M}$ or $\varphi_\mathrm{v}$ & $T_\mathrm{D}$ (K)\\
\hline
$M^\mathrm{osc}$ & 0.81(4) & 159.10(9) & 0.49(2)$\pi$ & 16.99(5)\\
$V_\mathrm{ME}^\mathrm{osc}$ & $6.40(7)\times10^{-4}$ & 157.05(2) & 0.88(1)$\pi$ & 8.88(4)\\
\hline
\hline

\end{tabularx}
\end{table*}

The observation of a single fundamental frequency in both magnetization and the magnetostrictive coefficient enables a direct comparison of their phases. The phases of quantum oscillations are commonly extracted either by directly fitting the oscillatory waveform to the LK expression or indirectly from the intercept of a Landau-level fan diagram. In the present field range, however, the lowest accessible Landau index for $F\sim160$ T remains above 10. Therefore, extrapolation to zero index would introduce substantial uncertainty. We thus determine the phases directly from LK fits. A similar strategy was used in the high-field magnetotransport study of \ce{YbMnBi2} to avoid ambiguities from multiple frequencies and harmonics \cite{Liu2017_YbMnBi}. Because the bulk Fermi surface retains finite $k_z$ dispersion, we use the 3D LK form with one fundamental frequency. At a fixed temperature $T$, the fitting functions can be written as:
\begin{equation}
    M^\mathrm{osc}= C_\mathrm{M}\frac{\exp{(-aT_\mathrm{D}\mu/B)}}{B^{1/2}\sinh{(aT\mu/B)}}\sin\left(\frac{2\pi F}{B}-\varphi_\mathrm{M}\right),
    \label{eq:M-fit}
\end{equation}
\begin{equation}
    V^\mathrm{osc}_\mathrm{ME}= C_\mathrm{V}\frac{\exp{(-aT_\mathrm{D}\mu/B)}}{B^{3/2}\sinh{(aT\mu/B)}}\sin\left(\frac{2\pi F}{B}-\varphi_\mathrm{V}\right),
    \label{eq:V-fit}
\end{equation}
where $a=2\pi^2k_\mathrm{B}m_0/(\hbar e)\simeq14.69$ T/K, $\mu=m^{*}/m_0$, $T_\mathrm{D}$ is the Dingle temperature, $\varphi_\mathrm{M}$ and $\varphi_\mathrm{V}$ are the corresponding phases in oscillating magnetization and ME signal, respectively. We adopt a common positive-amplitude convention for the field-independent parts of the two amplitudes ($C_\mathrm{M},C_\mathrm{V}>0$):
\begin{equation}
    C_\mathrm{M}\propto\left|\left(\frac{e}{2\pi\hbar}\right)^{3/2}\frac{2e\hbar aT\mu F\cos(\pi g\mu/2)}{\pi m^{*}\sqrt{|\partial^2 S_m/\partial k^2_\parallel|}}\right|,
    \label{eq:CM}
\end{equation}
\begin{equation}
    C_\mathrm{V}\propto\left|\left(\frac{e}{2\pi\hbar}\right)^{3/2}\frac{4e\hbar aT\mu kF^2\cos(\pi g\mu/2)}{m^{*}\sqrt{|\partial^2 S_m/\partial k^2_\parallel|}}\frac{\partial\ln F}{\partial\sigma_i}\frac{\partial E}{\partial\epsilon_i}\right|.
    \label{eq:CV}
\end{equation}
In practice, the effective $g$ factor, the strain-transfer coefficient $k$, the stress derivative $\partial S_m/\partial\sigma_i$, and the curvature $\partial^2S_m/\partial k_\parallel^2$ are not known. These field-independent quantities are therefore absorbed into $C_\mathrm{M}$ and $C_\mathrm{V}$. We fix $m^{*}=0.22m_0$ and $\mu=0.22$, while allowing the frequency to vary as an internal consistency check. Each fit thus contains four free parameters: $C_\mathrm{M}$ or $C_\mathrm{V}$, $F$, $T_\mathrm{D}$, and the phase $\varphi_\mathrm{M}$ or $\varphi_\mathrm{V}$.

As shown in Figs.~\ref{fig:2}(c) and \ref{fig:2}(f), the LK functions reproduce both oscillatory traces well. The fitted parameters are summarized in Table~\ref{table:1}, and the fitted frequencies agree with the FFT results. More importantly, under the common positive-amplitude convention, the phases in the oscillatory magnetization and ME voltage are $\varphi_\mathrm{M}=0.49\pi$ and $\varphi_\mathrm{V}=0.88\pi$, respectively. Their difference, $0.39\pi$, is reasonably close to the expected quadrature shift of $\pi/2$ [Eqs.~(\ref{eq:M-leading}) and (\ref{eq:lambda-leading})] given the restricted field window and slowly varying terms neglected in the leading-phase approximation. This phase shift is the central result of this work. Accordingly, a $\pm\pi/2$ correction, with its sign fixed by the stress derivative of extremal Fermi surface, must be applied before the cyclotron-orbit phase is inferred from the ac magnetostrictive coefficient.

The fitted phases should not, however, be identified directly with the Berry phase. The orbital-moment phase is not independently known, and an unknown sign of $R_\mathrm{S}$ or of $\partial F/\partial\sigma_i$ can introduce an additional phase shift of $\pi$ \cite{FOMINYKH_QO}. We therefore refrain from assigning a unique numerical Berry phase from the present phase comparison. Independent information about the orbital-moment contribution, effective $g$ factor, and stress response of the extremal orbit would be required for such an assignment. The Dirac-derived Fermi surface of \ce{YbMnBi2} is nevertheless consistent with the anomalous Hall and anomalous Nernst responses reported previously \cite{pan2022_YbMnBi,Guo_YbMnBi,Wen_YbMnBi}.

Finally, we examine how a small field misalignment affects the oscillation phase. Controlled rotation is difficult in both the VSM module and the homemade coil used for the magnetostrictive coefficient measurements. We therefore use thermopower $S_{xx}$, another sensitive quantum-oscillation probe, to determine the angular evolution of the frequency and phase. Figure~\ref{fig:3} shows $S_{xx}$ measured at 6.5 K for fields close to the crystallographic $c$ axis. The nominal angle $\theta$ is defined between the magnetic field and the $c$ axis [see inset of Fig.~\ref{fig:3}(d)]. Clear oscillations persist over $-10^{\circ}\leq\theta\leq15^{\circ}$, as displayed in Figs.~\ref{fig:3}(a) and \ref{fig:3}(b). Applying the same fitting protocol, we extract the angle-dependent frequency and phase of the oscillatory thermopower. The frequency varies from approximately 159.3 to 161 T over the measured angular range [Fig.~\ref{fig:3}(c)], making a slight difference in sample orientation a plausible explanation for the frequency mismatch between the magnetization and ME measurements. The minimum in the angular dependence further indicates a zero-angle offset of approximately $2.5^{\circ}$, arising from the unavoidable uncertainty in sample mounting and alignment. By contrast, the fitted phase remains nearly constant throughout the measured angular range [see Fig.~\ref{fig:3}(d)]. Importantly, no metamagnetic transition is observed in any thermopower trace measured up to 14 T over the investigated angular range, thereby ruling out a field orientation-induced magnetic reconstruction and the associated topological phase transition that could otherwise alter the phase of quantum oscillations. Thus, a misalignment of a few degrees does not measurably alter the phase of quantum oscillations in \ce{YbMnBi2}, supporting the reliability of the phase comparison between magnetization and the ac magnetostrictive coefficient.

\section{Conclusions}

In summary, we have compared the phases of quantum oscillations measured in magnetization and the ac magnetostrictive coefficient of the topological antiferromagnet \ce{YbMnBi2}. A single fundamental frequency near 160 T is resolved in both probes, enabling a direct phase comparison. Their oscillatory components exhibit an approximately $\pi/2$ relative phase shift, in agreement with the thermodynamic Maxwell relation connecting the magnetostrictive coefficient to the stress derivative of magnetization. Angle-dependent measurements of quantum oscillations in thermopower further show that field misalignments of a few degrees affect the oscillation frequency but do not produce a systematic phase shift. This calibration provides the basis for extracting cyclotron-orbit phase information from quantum oscillations measured through the ac magnetostrictive coefficient, provided that the additional orbital, spin, and stress-dependent phase factors are independently constrained.

\section*{Acknowledgements}

This work was supported by the National Key Research and Development Program of China (Grant No. 2025YFA1411301), the National Natural Science Foundation of China (Grant Nos. 12374081, 52401263, 12474142), the National Key R\& D Program of China (2025YFF0524500), the Scientific Research Innovation Capability Support Project for Young Faculty (Grant No. SRICSPYF-ZY2025076), the Open Projects at the Beijing National Center for Condensed Matter Physics (2025BNLCMPKF010), the Natural Science Foundation of Chongqing, China CSTC (Grant No. CSTB2024NSCQ-QCXMX0002), the Chinesisch-Deutsches Mobilit\"atsprogramm of the Chinesisch-Deutsches Zentrum f\"ur Wissenschaftsf\"orderung (Grant No. M-0496). 

\section*{Appendix: Theoretical background for quantum oscillations in magnetization and magnetostriction} \label{apa}
The oscillatory magnetization $M^\mathrm{osc}$ and magnetostrictive strain $\epsilon_i^\mathrm{osc}$ can be obtained from the oscillatory thermodynamic potential density $\Omega^\mathrm{osc}(B,\sigma_i)$ \cite{Chandrasekhar_MS}:
\begin{equation}
    M^\mathrm{osc}=-\left(\frac{\partial\Omega^\mathrm{osc}}{\partial B}\right)_{\sigma_i},
    \label{eq:thermo-M}
\end{equation}
\begin{equation}
    \epsilon^\mathrm{osc}_i=-\left(\frac{\partial\Omega^\mathrm{osc}}{\partial\sigma_i}\right)_{B},
    \label{eq:thermo-epsilon}
\end{equation}
where tensile stress is taken as positive. For a 3D extremal orbit, the LK expression in SI units is \cite{LK1956,Chandrasekhar_MS,Hu_QO}
\begin{equation}
    \Omega^\mathrm{osc}=\sum_{r=1}^\infty A_r\cos\left(\frac{2\pi rF}{B}-\Phi_r\right),
    \label{eq:Omega-LK}
\end{equation}
with
\begin{equation}
    A_r=\left(\frac{e}{2\pi\hbar}\right)^{3/2}
    \frac{e\hbar B^{5/2}}
    {\pi^2m^{*}r^{5/2}\sqrt{|\partial^2 S_m/\partial k^2_\parallel|}}
    R_{\mathrm{T}}R_{\mathrm{D}}R_{\mathrm{S}}.
    \label{eq:A-LK}
\end{equation}
Here $m^{*}$ is the cyclotron mass and $k_\parallel$ is the wave vector parallel to the magnetic field. The damping factors are
\begin{equation}
     R_\mathrm{T}=\frac{raT\mu/B}{\sinh(raT\mu/B)},
    \label{eq:RT}
\end{equation}
\begin{equation}
    R_\mathrm{D}=\exp{\left( -\frac{raT_\mathrm{D}\mu}{B}\right)},
    \label{eq:RD}
\end{equation}
\begin{equation}
    R_{\mathrm{S}}=\cos\left(\frac{r\pi g\mu}{2}\right),
    \label{eq:RS}
\end{equation}
where $a=2\pi^2k_\mathrm{B}m_0/(\hbar e)\simeq14.69$ T/K, $m_0$ is the free-electron mass, $\mu=m^{*}/m_0$, $T_\mathrm{D}$ is the Dingle temperature, and $g$ is the effective $g$ factor. The phase can be parameterized as $\Phi_r=2\pi r\gamma-2\pi\delta$, where $\delta=+1/8$ ($-1/8$) for the 3D minimum (maximum) orbit in the convention adopted here.

Substituting Eq.~(\ref{eq:Omega-LK}) into Eqs.~(\ref{eq:thermo-M}) and (\ref{eq:thermo-epsilon}) and retaining the derivative of the rapidly varying phase gives Eqs.~(\ref{eq:M-leading}) and (\ref{eq:epsilon-leading}). Here, the slowly varying terms $\partial A_r/\partial B$ and $\partial A_r/\partial\sigma_i$, together with any stress dependence of $\Phi_r$, are neglected. This leading-phase approximation is valid in the large quantum number regime, where $2\pi rF/B\gg1$ \cite{Chandrasekhar_MS}. Applying the same approximation when differentiating $\epsilon_i^\mathrm{osc}$ with respect to $B$ gives Eq.~(\ref{eq:lambda-leading}).

\clearpage
\nocite{*}

\bibliographystyle{apsrev4-2}

\end{document}